\documentstyle[twocolumn,aps,epsfig]{revtex}
\begin{document} 
%
{\bf Comment on
     ``Possible Spin Polarization in a One-Dimensional Electron Gas"}

In a 1996 letter K.J. Thomas {\it et al.} report on the discovery
of a conductance anomaly at $0.7\,(2e^2/h)$ observed in quantum
ballistic transport in split-gate quantum point contacts
\cite{Thomas96}.
Independently, Tscheuschner and Wieck observed
a similar structure at $0.5\,(2e^2/h)$ in quantum ballistic transport
in focused-ion-beam written in-plane-gate transistors
\cite{Tsch96}.
Actually, both observations were presented at NANOMES 96
in May 96. However, signals showing this type of structure
were recorded ealier, but passed uncommented on so far
\cite{vanWees91,Frost93}.
Indeed, the structure was already recorded in measurements
involving focused-ion-beam written in-plane-gate transistors
as well
\cite{deVries94}.

It should be emphasized that the observations of both groups
are fully compatible.
Whereas the Cavendish group scanned a temperature range
from 70 mK to 1.5 K observing
a weakening in the definition of the quatization
for higher temperatures,
the team at Bochum, due to experimental limitations,
was only able to perform experiments in the 1.3 - 4 K temperature range.
Both groups attribute the new effect to a manifestation of spin-polarized
transport, the polarization being of spontaneous nature here. The crucial
test is an application of an in-plane magnetic field
which makes the $0.7\,(2e^2/h)$ structure discovered by K.J. Thomas
{\it et al.\/} {\it continuously\/} approach $0.5\,(2e^2/h)$
for high magnetic field strengths \cite{Thomas96}.
This is in correspondence to the behavior seen
in in-plane-gate transistors
(cf.\ Fig 7 of Ref.\
\cite{Tsch96}),
where an external magnetic field in transport direction
{\it stabilizes\/} the $0.5\,(2e^2/h)$ structure.
The new quasi-plateau is comparatively robust 
in that it remains stable
even when all the quantized plateaux are washed out,
e.g.\ for high temperatures as well as high density
of impurities.
In other words, the new quasi-plateau is probably due
to a many-body effect showing some rigidity
similar to itinerant ferromagnetism
reflecting the idea of Gold and Calmels
\cite{Gold96}.
\par
Recent measurements in a different type of quantum wire
by Ramvall {\it et al.} observed anomalies at $0.2\,(2e^2/h)$
\cite{Ramvall97}.

\par
\centerline{\psfig{figure=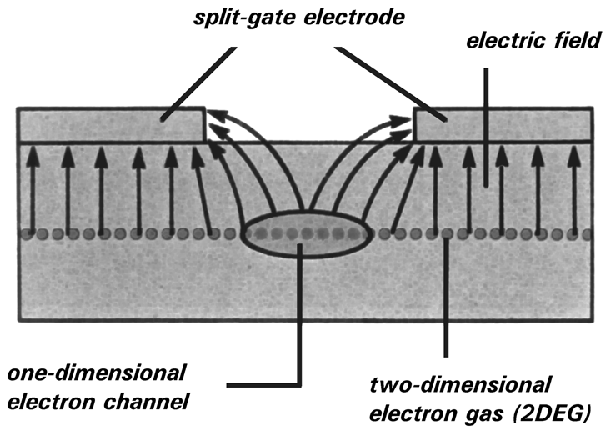,width=7cm,height=4cm}}
\centerline{{\bf Figure 1:} Split-gate set-up}
\centerline{$\phantom{X}$}
\par\noindent%

It is the purpose of this comment to discuss the question whether
the observed non-integer quantization by K.J. Thomas {\it et al.}
\cite{Thomas96},
which was always found at $0.7\,(2e^2/h)$, is universal or,
rather, related to the physics of the set-up.
We propose that, despite its
very special pecularities
\cite{Melngailis87,Orloff93}
and serious flaws such as
wide lateral spreading
of the implanted ions
\cite{Bever92},
which spoils the quality of quantization at low temperatures
\cite{Tsch96},
energetically carefully tuned
focused ion beam lithography
could shed some light on this
fundamental question.
In addition, to test the magnetic character of this many body effect
an obvious strategy would be to use the focused-ion-beam set-up to
implant a few magnetic ions such as Mn, Er, Yb.

Essentially, there may be three main points to be discussed
depending on the interpretation of the factor 0.7, namely as
fractional quantization, as orientation polarization,
or as an impurity effect.

The first interpretation suggests an interpretation in terms of
a fractionalization of charge analogous to the fractional
quantum Hall effect. There is an interesting result 
by Alekseev, Cheianov, and Fr\"ohlich
\cite{Alekseev97},
who, starting from the ideas of Landauer and B\"uttiker
and adapting ideas of current algebra, analyze in detail
how the system in question can be coupled to external reservoirs
determining the renormalization of the quantized conductance.
In particular, their parallel treatment of the quantum wire
and the fractional Hall effect problem reveals that, due to the
fact that physical electrons are to be identified with different
types of excitations, the wire does not allow for fractional
quantization in sharp contrast to the FQHE system. However,
a thorough quantum-field-theoretical treatment
of the quantum wire should include the spin-statistics
of the interacting electrons giving rise to
a {\it transported-spin-basis gauge structure\/}
\cite{Berry97}
not present in Luttinger-type models
which enforces a certain kind of {\it non-abelian\/} bosonization.
In this sense it is not clear at all how the elementary
excitations of the wire are related to physical electrons.
Hence, the possibility of fractional quantization remains,
though it is improbable.

The second interpretation points towards an oblique
orientation polarization of a current of magnetic dipoles
which is supported by the field distribution
in split-gate set-up as compared to the idealized in-plane-gate
set-up (see Figures 1+2).

On the one hand an obliquely polarized
state should approach a longitudinally or in-plane polarized
state smoothly under the influence of an increasing in-plane
external magnetic field which is in accordance with the
experimental observations
\cite{Thomas96,Tsch96}.
On the other hand a current of obliquely polarized magnetic dipoles
naturally induces an effective Hall-type voltage perpendicular
both to transport and polarization direction
which, as a part of electrodynamic response,
tends to compensate the oblique electric field components
induced by the top split gate geometry.
In an fairly idealized situation, in-plane-gates produces only
field components which are in-plane, such that the only
symmetry axis of the problem is the transport direction.

Note that the oblique-polarization picture is intimately related
to the semi-relativistic spin-orbit interaction picture.
More precisely, in electrodynamics the force on a moving
magnetic dipole in an external electrostatic field is
{\it dual\/} to the Lorentz force on a charged
particle in an external magnetic field in the sense
of electro-magnetic duality.
For example, if the Hall effect expresses
a balance between a tranverse electromotive force and
a Lorentz force, the appearance of an induced voltage
transverse to a plane, in which {\it off-plane\/} oriented
magnetic dipoles or vortices move, balances the spin-orbit
interaction. Hence, the latter depends on the geometry of
the contact. In fact, for a symmetric idealized
\lq\lq open X\rq\rq\ in-plane-gate transistor
we would have no spin-orbit interaction in contrast
so a set-up using sufficient asymmetric top gates,
which break inversion symmetry.
However, in our view the primary {\it in-plane\/} effect
is always due to spontaneous polarization,
which exhibits some rigidity which cannot be
explained by spin-orbit interaction.

\par
\centerline{\psfig{figure=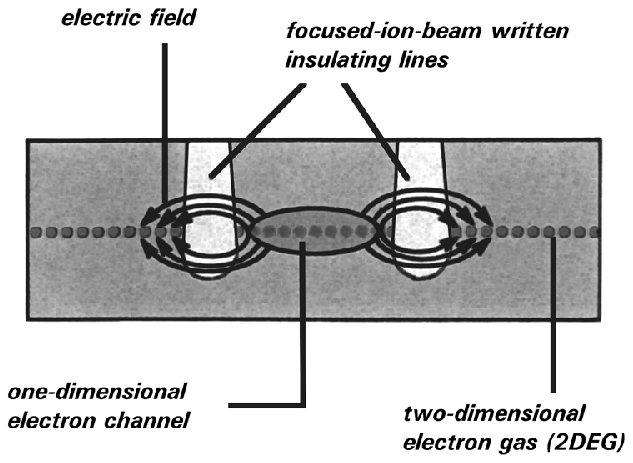,width=7cm,height=4cm}}
\centerline{{\bf Figure 2:} In-plane-gate set-up}
\centerline{$\phantom{X}$}
\par\noindent%

Unfortunately, we do not know at all to what extent real-world
focused-ion-beam written in-plane-gate transistors
reflect the pretension of its inventors.
To date, there is no theory of the density of
states of the focused ion beam implanted lines whose microscopic
structure is fairly fractal and dirty, in the mesoscopic regime
\lq\lq open X\rq\rq\ and \lq\lq open T\rq\rq\ are not too
different. It is a careful variation of the implantation energy
which could provide us with information about this point and should
have some influence of the quantization parameter.

The third interpretation seems to be the most natural.
As shown by several authors impurities in a narrow channel
alter the quality of the quantization of the conductance
\cite{Haanappel89,Marel89,Chu89}.
Chu and Sorbello showed that for $s$-like scatterers
the value is lowered
\cite{Chu89},
Haanappel and Marel studied the deformation
of the quantization staircase function
within the framework of quantum mechanical model calculations
\cite{Marel89,Haanappel89}.
In Ref.\
\cite{Tsch96}
the strong deviation of the standard $(2h/e^2)$
plateau was attributed to this effect, however,
the new state seems to be more stable against the
influence of impurities. This is in harmony with
the work of K.J. Thomas {\it et al.\/}.
Future study should focus on the different behavior
of the impurity dip for the conventional and the
novel plateaux. In particular, the magnetic character
of this many-body effect could be traced by
implanting a few magnetic (Kondo-type) impurities.

In conclusion,
it cannot be overemphasized that
unlike standard ballistic quantization
the effect in question shows up clearly
not only in ultra-clean long quantum wires
(find the tiny sharp bend at $0.6\,(2e^2/h)$
 in Fig.\ 1 at  of Ref.\ \cite{Tarucha95})
but also in \lq quick and dirty\rq\ prepared systems
such as focused-ion-beam
written in-plane-gate contacts.
This is reminiscent of the robustness of superconductivity
and other many-body phenomena.
To study the fundamental nature of the value
of the new quantization step it is proposed
to vary carefully the implantation energy.
Finally we remark, that would be interesting
to implant a few {\it magnetic ions\/} to test
the many-body character of this magneto-electronic
mesoscopic effect.

\vskip .2cm
\vskip .1cm
%
{\small
R.D. Tscheuschner and T. W\"olkhausen
\vskip .08cm
\indent
%
%
I.\ Institut f\"ur Experimentalphysik der Universit\"at Hamburg,
Luruper Chaussee 149, D-22761 Hamburg, Germany
%
%
\vskip .2cm
\noindent Received
\vskip .2cm
\noindent
The authors would like M. Dinter and
R.L. Stuller for discussions.
One of us (R.D.T.) would like to thank
H. Schulz and K.J. Thomas
for email correspondence.
\vskip -.6cm
%
%
}
\end{document}